
\magnification 1200
\def\rf{\hfill\break}
\def\etal{{\it et al. }}
\def\l{\lambda}

\def\pt{\  .}
\def\com{\  ,}

\def\g{\gamma}

\def\l{\lambda}

\def\md{\dot M}

\def\sles{\lower2pt\hbox{$\ \buildrel {\scriptstyle <}
   \over {\scriptstyle\sim}\ $}}
\def\sgreat{\lower2pt\hbox{$\ \buildrel {\scriptstyle >}
   \over {\scriptstyle\sim}\ $}}
\bigskip
\rf
\centerline {\bf COSMIC RAYS FROM ACCRETING }
\centerline{\bf  ISOLATED NEUTRON STARS}
\bigskip
\bigskip
\centerline {AMOTZ SHEMI}
\bigskip
\centerline{ Wise Observatory \& School of Physics and Astronomy}
\centerline{ Raymond and Beverly Sackler Faculty of Exact Sciences}
\centerline{ Tel Aviv University, Tel Aviv, 69978, Israel.}
\centerline{ e-mail:amotz@wise.tau.ac.il}
\vskip 3cm
\rf
{\it To be published in MNRAS}
\bigskip
\rf
{\bf ABSTRACT}
\medskip
\rf
Interstellar matter that is accreted onto isolated magnetic neutron
stars in the Galaxy ($\sim 10^9$ by number) is accelerated and
reflected back by MHD shocks, which envelope the stars. The integrated
power in the Galaxy $L_{cr,ns}$ is $ \sgreat 10^{40}  {\rm erg \ s^{-1}
}$, the energy distribution is a power law of spectral index $> 2$, and
the particle energy can be raised to $10^6$ GeV, consistent with the
power and spectrum of primary cosmic rays in the Galaxy.

The major
contribution for $L_{cr,ns}$ comes from a minority of $\sim 10^7$
isolated neutron stars which are located within dense clouds.
Sources in these clouds, that are generally spread within the Galactic
disk, can explain the concentration of high-energy cosmic rays in the
Galactic plane, as deduced from pion decay spectra in gamma-ray
observations.
The soft X-ray luminosity from these neutron stars
is consistent with the Galactic X-ray background.
The accretion may be associated with
ion-neutral bias, that is further enhanced by ion confinement in
frozen-in magnetic fields,
which can raise the relative abundance of first ionization potential (FIP)
elements in the cosmic rays.

\bigskip
\rf
\vfill\break
\rf
{\bf 1 INTRODUCTION}
\medskip
\rf
The origin of cosmic rays has long been a central question in astrophysics
(e.g. Cesarsky 1987, hereafter C87, Blandford \& Eichler 1987 [BE87],
Axford, 1994).
The angular distribution of cosmic rays is highly
isotropic. The isotropy accuracy is $\sim 0.1\%$ at energies
$\sles 10^4$ GeV and $\sim 1\%$ at $10^{8}$ GeV, but
it is limited to a few tens of per cent in the
ultra-high-energy range
$ > 10^{9}$ GeV. The origin of cosmic rays,
apart of the ultra-high-energy class,
is believed to be Galactic, associated with continuous
leakage from the Galactic magnetic field confinement.
The confinement time is observationally
constrained by the decay time of unstable nuclei,
e.g. $^{10}$Be ($t_d = 2.2 \times 10^{6}$ yr), or
$^{26}$Al ($t_d = 0.85\times 10^{6}$ yr), to be $\sim 8\times 10^6$ yr
(for homogeneous ISM).

Three major requirements of candidates for cosmic ray sources are
that they produce the observed cosmic ray power,
spectra, and elemental abundance. The total
power of the Galactic component is estimated to be
$ L_{cr} \sim 2 - 3 \times 10^{40} \ {\rm erg \ s^{-1}}$ (BE87).
The spectra, which are noisy and depend on the specific ion,
are approximately a power low $ \propto E^{-\alpha}$,
where $ \alpha < 2.5 $ at energies below $ \sim 10^2$ GeV, $\sim 2.7$
up to $10^{6.5}$ GeV and $\sim 3$ at $\gg 10^{6.5}$ GeV.
The elemental composition (Meyer 1985a,b, Silberberg \etal 1987)
at energies up to $10^6$ GeV clearly deviates
from the Local Galactic abundance,
but fairly agrees with the Coronal composition.
Specifically, there is a relative enhancement in cosmic rays of
the low FIP ($<10$ eV) elements.

Supernova remnants (SNRs) that heat and accelerate the surrounding material
via shock wave acceleration are widely accepted as
the origin of Galactic cosmic rays
(Bell 1978, Blandford \& Ostriker 1980, BE87).
The cosmic ray power is only a few per cent of the
total power associated with SNR mass motions, therefore, even an inefficient
shock acceleration mechanism is applicable. Such a mechanism also
explains the spectral shape.
The maximal energy gain per
particle $E_{max}$, however, is limited by the shock lifetime, estimated to be
$\sim 10^5 B/(10^{-6} {\rm G})$GeV in optimal conditions ($E_{max}$
may be increased to $5\times 10^6$ GeV if
the SNR is propagating into a stellar wind cavity [V\"olk \& Biermann 1988]).
Moreover, the FIP bias is unexpected, since
the SNR is accelerating material which is believed to have
the Local Galactic abundance
(this problem is relaxed in models of SNR in a stellar wind cavity,
if the presupernova object is a red or a blue giant, or a
Wolf-Rayet star [Silberberg \etal 1990]).
In competing 'stellar' models sub-relativistic particles are injected
from F and G stars (Meyer 1985a,b), which explain the cosmic ray composition,
but
require a secondary stage of acceleration, e.g. by passing a supernova shock
wave.
An injection mechanism for ultra-high-energy cosmic rays was
proposed by Kazanas \& Ellison (1986), based on
accreting binary systems such as massive X-ray binaries, where
a strong stellar wind of the companion star is accreted onto
the compact object,
a neutron star (NS) or a black hole.

Cowsik \& Lee (1982) have shown that accretion flows with shocks onto
isolated NSs can accelerate cosmic rays very efficiently.  They
consider old NSs, assuming that their surface magnetic fields are weak
and do not affect the hydrodynamics. The flow stagnation imposed by the
stellar surface results in a standing spherical shock wave which
envelops the star very close to its surface.  Potential energy,
associated with interstellar matter (ISM) accreted onto the star, is
continuously channeled to acceleration of high energy particles. The
cosmic ray power is $\sim 10^{36} \ {\rm erg \ s^{-1} }$ per NS,
assuming that its velocity is smaller than the ISM sound speed, and the
ISM density is large $\sim 10^3 {\rm \ cm^{-3}}$. For such a luminosity
a few times $10^4$ NSs are sufficient to explain the Galactic cosmic
rays.  Such a high power, $\sim 10^{-2}$ of the Eddington luminosity,
may be, however, associated with a temperature increase above the pair
production threshold, as was recently shown by Turolla \etal (1994).
Pair annihilation and gamma ray emission can significantly cool the
particles and (electron) synchrotron would become very efficient,
whether the magnetic fields are originated in the NS or
frozen within the accreted ISM.
Also, if the peculiar velocity of an old NS
is larger than the ISM sound speed the available power is significantly
smaller than $10^{36}{\rm \ erg \ s^{-1}}$.  More importantly,
various observations suggest that the magnetic field of an old NS
remains very strong (e.g. Srinivasan 1989, Phinney \& Kulkarni, 1994),
and in that case the magnetic pressure can halt the flow at the Alfv\'en
surface far from the stellar surface.

In this paper I focus on Galactic cosmic rays that originate from
accreting magnetic isolated NSs.  Because of the halting of the
accretion, a standing shock is created, within which shock acceleration
occurs (Shemi 1995). The available power, compared with the free fall
luminosity,
is significantly reduced, to a value orders of magnitude below the Eddington
luminosity.
Other mechanisms that might regulate the accretion, such as the "propeller
effect"
(Illarionov \& Sunyaev 1975) or a relativistic wind induced by the NS,
which will not be discussed here.
Accretion through the polar caps, which, unlike disk-accretion,
probably is inefficient in spherical inflows, is ignored, and
instabilities of the magnetosphere are not considered. The
accreted matter is never heated to ultrarelativistic temperature, and
pair production therefore does not take place.

The total population of Galactic isolated NSs is estimated to be $N \sim 10^9$
(Blaes \& Madau 1993, hereafter BM93), but
the major contribution
to cosmic rays comes from a small subgroup ($\sim 1\%$)
containing the slow NSs which are located within dense clouds.
In the following paper I first describe the physics of
accretion, shock waves and particle acceleration around an
isolated magnetic NS, and
then discuss the applicability of these objects as sources of cosmic rays.

\bigskip
\rf
{\bf 2 ACCRETION ONTO A MAGNETIZED NEUTRON STAR }

\rf
An isolated NS that is propagating with velocity $v_{40} = v/(40 \
{\rm km \ s^{-1} })$ accretes ISM
from a cylinder of a Bondi-Hoyle radius
$$
r_{BH} = {2 G {\rm M} \over v^2 + c_s ^2 } \simeq
2.3 \times 10^{13} v_{40}^{-2} \ {\rm cm} \pt
\eqno(1)
$$
Here $c_s$ is the sound speed and M is the NS mass, where throughout
this paper we use M $= 1.4 {\rm M}_{\odot}$. I will focus on dense ISM
clouds ($n  =  10^4n_4 \ {\rm cm^{-3} }$), where the accretion rate
would be large.  Extreme UV and soft X-ray radiation emitted by the
accreted plasma is sufficiently intense to ionize the matter within a
Str\"omgren sphere $ r_S\sim 3 \times 10^{15} (S/10^{43}) n_4^{-2/3}
(\alpha_B/10^{-12})^{-1/3} \ {\rm cm}$, where $S$ is the rate of
ionizing photons, in photons s$^{-1}$, and $\alpha_B$ is the
recombination coefficient to excited states of hydrogen, in  ${\rm cm^3
\ s^{-1}}$. In the ionized gas macroscopic magnetic fields and
collective plasma phenomena couple the particles, and the accretion
become hydrodynamical, rather than collisionless.  Removal of angular
momentum becomes efficient and raises the accretion efficiency, by a
factor as large as $(c/c_s)^2 \sim 10^9$, compared with that of a
neutral gas (e.g. Shapiro \& Teukolsky 1983).

Far from the central object ($ r_* \ll r \sles r_{BH}$, where
$r_* $ is the NS radius, $10^6$ cm in this paper)
irreversible
processes of energy dissipation are negligible,
and the pressure obeys the adiabatic relation
$$
p = p_0 \Bigl( {\rho \over \rho_0}\Bigr)^{\Gamma} \com
\eqno(2)
$$
where $\Gamma$ is the specific heat ratio.
The expected accretion rate is
$$
\md = 4 \pi r_{BH}^2 \lambda_s
\rho v \simeq 4.3 \times 10^{14} n_4 v_{40}^{-3} {\rm g \ s^{-1} } \com
\eqno(3)
$$
where $\lambda_s$ is a numerical factor of order of unity.  The accretion
rate is much smaller than the Eddington rate $\dot M_{Ed} = L_{Ed} r_*
/(G{\rm M}) \sim 10^{18} \ {\rm g \ s^{-1} }$, therefore radiation pressure
is never strong enough to halt the accretion.

The continuous removal of angular momentum causes streamlines to bend
and converge toward the NS. To follow the flow dynamics we neglect
the influence of magnetic fields and back streams from the central object
or from the shock, and assume spherical symmetry.  The dynamics are
governed by the continuity equation (3) and by the Bernoulli equation
$$
{v^2 \over 2} + {c_s^2 \over \Gamma -1} - {{\rm GM} \over r} = 0 \pt
\eqno(4)
$$
The sound speed [$c_s = (\g kT/m)^{-1/2} =
0.57  \ (T/50 {\rm\ K^o)^{1/2} \  km \ s^{-1} }$]
of the dense and cold ISM cloud is low,
therefore, even for small NS velocity,
the flow at $r = r_{BH}$ is supersonic, with a rather large Mach number
$\bar M  = 70\ v_{40} T_{50}^{-1/2} $.
Unless nonthermal acceleration occurs,
the energy gain per particle does not exceed
$[1 - (v_{ff}/c)^2]^{-1/2} - 1 = 0.34 $ of its rest mass,
maintaining the plasma  subrelativistic even when approaching the NS, with
typical blackbody temperature $\sim 10^6 - 10^7 \ {\rm K}$.
Substituting the nonrelativistic value $\Gamma = 5/3$ and the limit
$c_s \ll v $, equations 2-4 give the scaling behavior of $v, \ c_s, \
\rho $, and the gas ram pressure $ \rho v^2$:
$$
v \propto r^{-1/2} \ ; \   c_s \propto r^{-1/2} \ ; \
\rho \propto r^{-3/2}; \ \rho v^2 \propto r^{-5/2} \pt
\eqno(5)
$$
Note that the Mach number remains constant.  The ionized gas has high
conductivity and the magnetic Reynolds number is large, so that the
residual magnetic fields in the ISM are frozen within the inflow.
Field
lines are swept along the streamlines and the global field pattern becomes
radial [one may consider a large - scale field which initially is homogeneous
$(B_x,B_y,B_z) = (B_0,0,0)$, in where
an accreting object is placed.
The lines of the frozen-in field would be bent towards the
central object, along the $y$ and $z$ axis,
resulting in a configuration
of a nozzle. The nozzle configuration is narrow in its central part and
globally the field lines surrounding the central object become radial].
The field strength
increases $\propto r^{-2}$ and the magnetic pressure would
therefore increase as $r^{-4}$, faster than the ram pressure and the
gas pressure ($\Gamma^{-1}\rho c_s^2 \propto r^{-5/2}$).
Such a field
enhancement would, however, be regulated when small scale
magnetic reconnections become sufficiently fast.
The field strength is therefore bounded by the equipartition
limit, that is achieved when the reconnection time scale, $r/v_{A}$,
becomes comparable to
the compression time scale, $r / v_{ff}$, $B_{gas}^2/(4 \pi)^{-1} \sles
\rho v^2$.  This occurs at the equipartition radius
$$
r_{eq} = 7 \times 10^{8}\bigl({B_{ISM} \over 3\times 10^6
{\rm \ G} }\bigr)^{4/3}
n_4^{-2/3} v_{40}^{-10/3} {\rm \ cm}
\eqno(6)
$$
where $B_{ISM}$ is the magnetic field strength at $r_{BH}$.

Magnetic fields of the strength expected
in old NS, $10^9 \ \sles \ B_{\rm surface} \  \sles 10^{12} {\rm G} $
(e.g. Srinivasan 1989, Phinney \& Kulkarni 1994), would cause stagnation
of the flow at the Alfv\'en surface, where the magnetic pressure balances
the flow ram pressure.  The flow is decelerated and compressed, and standing
shock waves occur ahead of the Alfv\'en surface.

Using the dipole approximation
$B \propto r^{-3}$, the Alfv\'en radius is given by
$$
r_m = 10^8 B_{10}^{4/7} n_{4}^{-2/7}v_{40}^{6/7}
\ {\rm cm} \pt
\eqno(7)
$$
Here $B_{10}$ is the NS surface magnetic field in units of $ 10^{10}$ G.

If the light corotation radius $r_{lc} = c P /2\pi = 4.8 \times 10^{9}
P_1 {\rm s} $ ($P_1 \equiv P/1 \ {\rm s}$) is larger then $r_m$ a
pulsar wind beyond the light cylinder can halt the inflow.  However, such
a mechanism would be significant if the NS rotation period is shorter
than
$$
P_{lc,m} = 2\times 10^{-2} B_{10}^{4/7} n_{4}^{-2/7} v_{40}^{6/7} \ {\rm s}
\com
\eqno(8)
$$
while old NSs, unless recycled, are generally slower than
$ \sim 1 $ s.
Beyond $r_{lc}$ the magnetic field decreases as $ r^{-1}$, therefore in
very fast rotators ($P < P_{lc,m}$) the Alfv\'en surface would
generally be pushed beyond $r_{BH}$.  If $r_m > r_{BH}$, the accretion
is prevented in the first place.  The motion of the NS through the ISM
is then associated with a stationary bow shock, distant
$$
r_{ism} = 3.2 \times 10^{13} B_{10} v_{40}^{-1} P_1^{-2} \ {\rm cm }
\eqno(9)
$$
ahead of the star. The accretion flow will overcome the centrifugal barrier
(where the free fall velocity become equal to the Keplerian velocity),
at the centrifugal radius $r_{cent} = (GM/\Omega^2)^{1/3}$
$= 1.7\times 10^{8}P_1^{2/3} {\rm \ cm, }$
if the star period
$P$ exceeds $ 2 \pi r_m^{3/2} (GM)^{-1/2} $.
With equation 7 this condition reads
$$
P > P_{cent,m} = 0.46 B_{10}^{6/7} n_4^{-3/7} v^{9/7} \ {\rm  s} \pt
\eqno(10)
$$
For NSs in dense ISM clouds we can reasonably assume $P > P{cent,m}$,
and therefore consider a shock wave at $r_m \sim 10^8/r_{m,8} {\rm \ cm}$.
\bigskip
\rf
{\bf 3 PARTICLE ACCELERATION}
\medskip
\rf
The topology and physics of the shock ahead of the Alfv\'en surface is
complex and not fully resolved yet.
Shock topology would differ from a spherical shape,
presumably is conical with an
opening angle $\sim \bar M ^{-1}$ and a bow front, and
the obliquity varies from parallel to perpendicular.
A large electron and ion concentration is built up in the foreshock
regions. Irregularities in the shock topology
and temporal variability are also expected, in the same manner as in
simulations of wind accretion onto an
{\it unmagnetized} massive object (Ruffert, 1994 and references therein).
Backward streams and perturbations influence the topology, sometimes
depending on the Mach number and the $r_{BH}/r_*$ ratio, making the
flow configuration unstable.

The jump conditions for a planar-discontinuity, infinitesimally narrow
shock imply a compression ratio $\zeta = \rho_2/ \rho_1$ (here '1' denotes
the upstream region and '2' stands for downstream)
that approaches the limiting value
$\zeta \rightarrow
(\Gamma + 1) /( \Gamma - 1) = 4 $ for $\Gamma = 5/3$
(e.g. Draine \& Mckee 1993).
The parallel component of
the magnetic field is conserved ($B_{1,\|} = B_{2,\|}$), while the
perpendicular component downstream is enhanced
($B_{2,\perp} = \zeta B_{1,\perp} $ for a flow with no
transverse velocity component
on both sides). The true strength of the magnetic fields upstream is reduced if
the NS dipole fields are screened by electric field ($\bar v\times \bar B$)
in the shock region. On the other hand, the
enhancement of frozen magnetic
fields in the inflow would increase the upstream field.

In the following I will neglect the complexity in the shock conditions, and
assume that known acceleration
mechanisms in planar shocks are still applicable, to assess the cosmic ray
luminosity, the maximal energy per particle and the energy distribution.

Some energy loss mechanisms take place at the shock region, among them
radiation, dissipation through MHD waves and escape of neutral particles.
Defining $ f_{cr} $ the fraction of the free fall luminosity
($l_{ff} =  \dot M G {\rm M} r_m^{-1}$)
that is left in cosmic rays after such losses,
we obtain the power per an accreting NS
$$
l_{cr,ns} = f_{cr} \dot M {G {\rm M} \over r_m}  =
8 \times 10^{32} f_{cr} \lambda_s B_{10}^{-4/7} n_4^{9/7} v_{40}^{-15/7}
  \ {\rm erg \ s^{-1} } \pt
\eqno(11)
$$

Particles would be accelerated through multiple shock crossing,
presumably via a {\it diffussive shock} first-order Fermi mechanism
(Drury 1983, BE87).  {\it Shock drift} acceleration, moreover, can take
place in the quasi perpendicular segments of the shock, where an
electric field $\bar v \times \bar B$ exists.  The nuclear mean free
path $l_{col} =  \l_{pp}/nm_p$ (where $\l_{pp} \sim 60 {\rm g
\ cm^{-2}}$) at the shock region is $\sim 3 \times 10^{13} n_4^{-1}
{\rm \ cm}$, much larger than the shock scale $r_m$.  However, the
proton Larmor radius $l_B  =   \gamma m_p c^2 /(eB)$ will be smaller
than $r_m$ unless particles are ultra relativistic ($\g \gg 10^5$).
Particle -  wave interactions will therefore dominate the inelastic
(particle - particle) collisions, making the shock collisionless.
Particles are scattered by Alfv\'en waves, small-scale magnetic
irregularities and turbulences, or other collective phenomena.

If the upstream magnetic pressure $P_{mag}$ is well below the gas
pressure $P_{gas}$, the MHD
perturbations are basically Alfv\'en waves that travel upstream with
velocity $v_A < v_1$. In the downstream region the plasma experiences
strong magnetic turbulences.  Particles are reflected by these
turbulences backward to the upstream region, then reflected again
downstream by the Alfv\'en waves (or also by turbulences, if
$P_{mag}\sim P_{gas}$ upstream.  Note that even in that case we expect a
large pressure gradient which would force a shock).  For a particle
with velocity $w$ the momentum gain per single shock crossing is $p
\rightarrow p\times 4[v_2-v_1]/3w$.  After a sequence of $\sim c/w$ random
crossings the momentum is raised to the relativistic regime, and, since
$w \gg v_A$, the Alfv\'en waves upstream are seen as a quasi-static
barrier.

When the acceleration is sufficiently rapid, and not affected
by energy losses,
the particle energy is raised
until the particles diffuse from the shock.
The conical shape of the shock allows
downstream accelerated particles to escape.
The diffusion coefficient may be assumed to have the general form
$\kappa = \lambda w/3$, where
$\lambda$ is the particle mean free path, usually between one to a few
Larmor radii.
Particles remain confined within the shock as long as the upstream diffusion
length scale $\kappa/ v_1$ is larger than the shock
radius of curvature $\sim r_m $.
The maximal energy per particle can be estimated by using
the Bohm approximation ($\lambda = l_B$) and
requiring $ (l_B w)/(3 v_1) < r_m$, that gives (with $w = c$;
$v_1 = (2 G M)^{1/2} r_m^{-1/2}$)
$$
E_{max} \equiv \gamma_{max} m c^2 \sim
5.4 \times 10^4 r_{m,8}^{-5/2}
B_{10} {\rm \ GeV} \pt
\eqno(12)
$$
To obtain a value of $E_{max}$ comparable to the 'knee' value $\sles
10^6$ GeV, at which the cosmic ray spectrum bends, one needs $r_{m,8} \sgreat
0.3$.  Since the Alfv\'en radius scales as
$r_m \propto B^{4/7}$, $E_{max}$ scales $ \propto
B^{-3/7}$, and such large
values of $E_{max}$ does not have to be associated with a large
accretion rate, if the stellar magnetic field is small $B_{10} \ll 1$.

Energy loss by synchrotron emission also can, in principle, regulate the
particle
acceleration. The maximal energy in that
case is determined by the competition
between these processes. The time scale for diffussive acceleration
$t_{acc} = p/\dot p \sim \kappa /v_1^2$ [more rigorously,
taking care for the time spent
by a particle in each side of the shock, one can obtain
$t_{acc} = 3\times (v_1 - v_2)^{-1}(\kappa_1/v_1 + \kappa_2/v_2$), e.g.  Drury
1983] while the ion cooling time scale $t_{synch,ion} \sim 6 \times
10^{18} A^3 Z^{-4}B^{-2}\gamma^{-1}$ s (here $B$ in Gauss). From  the
requirement $t_{synch,ion} > t_{acc}$ we obtain the constraint
$\gamma_{max,ion} \sles 2.6 \times 10^8 A^{3/2} Z^{-2} B_{10}^{-1/2}
r_{m,8} $ (note that in this case $\gamma_{max,ion}$ only weakly depends on
$B$,
namely $\gamma_{max} \propto B^{1/14})$.

The synchrotron cooling time of electrons is, however,
shorter then that of protons,
by a factor $(m_e/m_p)^{3}$, implying (with $\gamma_{max,e} \propto
(m_e/m_p)^{3/2}$)  $\gamma_{max,e} \sim 3.3 \times 10^3
B_{10}^{-1/2} r_{m,8}{\rm \ s}$. Synchrotron cooling
although negligible for ions below $\sim 10^8$ GeV, becomes
acute for electrons.
Moreover, electrons acceleration
is not even started unless additional 'injection' mechanism operates.
MHD waves upstream do not resonate
with electrons below $ \gamma_{min,e} - 1 = m_p/m_e \times (v_A/c)^2$ (for
$(c/v_A)^2 \gg (m_p/m_e)^2$, hence only the relativistic electrons
can be confined in the shock. The mechanism discussed here
therefore may not be applicable for cosmic ray electrons.

\medskip

The initial energy distribution of particles upstream is roughly a delta
function at $GM/r_m \sim (0.064 c)^2 r_{m,8}^{-1}$, with some thermal
broadening ($kT \sim c_s^2 \ll v_1^2$).  By the stochastic nature of the
acceleration, and since the total number of
scatterings per particle is sufficiently large, the distribution is
spread out with no preferred scale.
{}From the linear theory of acceleration in subrelativistic
shocks we expect a general form
$$
n(E) \propto E^{-\alpha}\pt
\eqno(13)
$$
The spectral index, given by
$$
\alpha = {\zeta + 2 \over \zeta - 1} \sim 2 \com
\eqno(14)
$$
is independent of the magnetic fields, and is found to be typical
in other cases of shocks.
Cowsik \& Lee (1982) show that a power law typical to cosmic rays can also be
obtained in shock waves which have a spherical geometry.
Ellison \& Eichler (1985) also show that $\alpha \sgreat 2$
in non-linear shock models that account for the cosmic ray feedback.
Even when the Alfv\'en Mach number gives rise to strong compression,
$\zeta \gg 4$, the slope is almost unaffected.

\bigskip
\rf
{\bf 4 DISCUSSION}
\medskip
\rf
The Galactic isolated NS population (BM93)
is estimated from the pulsar birthrate, assuming a steady state, to be
$N > 10^8$, but nucleosynthesis constraints on Galactic chemical evolution
require a total number
as large as $N\sim 10^9$ (Arnett, Schramm, \& Truran 1989).
A fraction of the potential energy associated with matter
accreted onto these NSs is continuously
channeled to accelerate particles to
very high energies.
Although the detailed shock structure and acceleration process
are yet unclear, the energy budget and the knowledge of
acceleration in astrophysical MHD shock waves
suggest that accreting isolated NSs are significant cosmic ray sources.
The cosmic ray luminosity $L_{cr,ns}$
is the integral over the cosmic ray emission $l_{cr,ns}$
from all the Galactic isolated NSs, and
other Galactic compact accretors like
isolated white dwarfs, black holes, and
massive binaries, that should be taken into account.

We briefly discuss the parameters that affect the accretion rate and
thereby the cosmic ray luminosity,
namely the ISM density distribution, the NS velocity distribution,
their surface magnetic fields and periods,
and the physics behind the efficiency factor $f_{cr}$.

The filling factor of the ISM dense phase
is estimated to be $\sim 1\%$ in the disk, and it rises
to $\sim 10\%$ in an extended region
in the Galactic center. From the estimated NS density in clouds
$ 5\times 10^{-4} (n_{ns}/{\rm pc^{-3}})(M_{cloud} /M_{\odot})^{3/2}$,
BM93 deduced $\sim 10^3$ isolated NSs in the 19 clouds closest ($\leq 800$ pc)
to the Solar System.

The rms value of their velocity distribution is calculated to be
$98\ {\rm km \ s^{-1}}$, where a substantial
fraction of them, $f_s \sim 0.25$, are slow ($v \sles 40 {\rm km \ s^{-1}}$),
and $f_d \sles 0.2 $ of them are located within dense ISM clouds
($n \sim 10^2 - 10^8 {\rm cm^{-3}}$). The upper density value
represents the cores of giant molecular clouds.
BM93 show a disk velocity distribution with
$6\%$ of NSs having $v \leq 20 \
{\rm km \ s^{-1}}$,
$22 \% $ below $40 \ {\rm km \ s^{-1}}$, and $50\%$ below
$72 {\rm \ km \ s^{-1} }$. It turns out that
the major contribution to $L_{cr,ns}$
would come from a minority of $10^7$
isolated NSs which are slow and located within dense clouds.

The strengths of magnetic fields of old NS are controversial. Field decay
is predicted in many
NS models, which have also attempted to explain why such decay slows down or
stops. However, there
are arguments against a decay mechanism, in which case the diversity in pulsar
magnetic fields is the result of
the conditions during their creation
(see e.g. Srinivasan, 1989 Phinney \& Kulkarni 1994).

The fraction $f_{cr}$ of the input power that is channeled to cosmic rays
depends on the efficiency of competing energy loss mechanisms.
Synchrotron radiation can maintain the electron temperature low
but only loosely affect the ions.
Free - free emission
[$\epsilon_{ff} \times r_m^3 \sim 10^{27} n_4^2 (T/10^7 {\rm\ K})^{1/2} \
{\rm erg \ s^{-1}}$] and Compton losses are also small.
Escape of
neutrons, which are produced through $pp \rightarrow nX$ reactions
or photodissociation of $^4$He nuclei, can
significantly ($\sles 25\%$)
affect the ultra-high-energy part of the spectrum
(Kazanas \& Ellison 1986). Energy is also dissipated through MHD
waves.
For the sake of
evaluating $L_{cr,ns}$ we will use $f_{cr} \sim 0.25$ and
focus on the slow stars in dense clouds.
Using the canonical values discussed above we have
$$
L_{cr,ns} = 1.2 \times 10^{40}
\bigr({N   \over 10^9}\bigl)
\bigr({f_s \over 0.25}\bigl)
\bigr({f_d \over 0.2}\bigl)
\bigr({ f_{cr}\over 0.25}\bigl)
\bigr({L_{ff} \over 10^{33}}\bigl)\ {\rm erg \ s^{-1} },
\eqno(15)
$$
which is consistent with the power estimated for Galactic cosmic rays.

The maximal energy gain per particle (Equation 12) is sufficient to
explain the majority of cosmic ray particles, but not the
very-high-energy ones.  With a large accretion rate $E_{max}$ would
greatly be raised, to values similar to those proposed by Kazanas \& Ellison
(1986) for binary systems.  Moreover, coherent acceleration through
magnetic field annihilation (frozen-in fields + NS fields) as is
suggested here, as well as pulsar acceleration, as proposed elsewhere
by many authors (e.g. Srinivasan 1989, Arons \& Tavani 1994), are
likely to take place.

Of particular interest is the subgroup of very slow ($v \sles c_s$)
isolated NSs in dense and cold clouds $ c_s < 1 \ {\rm km \ s^{-1}}$.
Although these strong accretors emit a large X-ray luminosity,
their Bondi - Hoyle radius $ \gg 10^{16} \ {\rm cm}$ can exceed the
Str\"omgren radius. The {\it effective} accretion radius, where
the gas is coupled through magnetic fields and angular momentum is
removed by fast collisions, shrinks
below $r_{BH}$.
Accreting NSs with $r_S < r_{BH}$ may enlarge
the cosmic ray (low FIP)/(high FIP) elemental ratio, that is observed to be
$\sim 3$ times larger than
in Local Galactic matter (Meyer 1985).
Selection between ions and neutral atoms occurs
far from the NS, at the radius
where each element becomes ionized. The radii of photoinization fronts
would increase as the ionization potential decreases (note that the
ionized continuum is much flatter than blackbody).
By assumption, ions are more effectively accreted than neutrals,
and by such an ion - neutral bias
the relative fraction of low FIP elements in the inflow is enhanced.
Furthermore the ions would be anchored within the inflow by the frozen
magnetic fields, while neutral atoms
can diffuse in a direction oppose to the temperature gradient,
As long as the diffusion time scale $r^2/D$ is smaller then the
dynamic time scale, $r/v$, ion-neutral separation by diffusion
can be significant.
The diffusion coefficient for neutrals in an ionized hydrogen gas (Geiss 1982)
$D =1.1 \times 10^{21} T/(10^4 \ {\rm K})^{3/2}
n_1^{-1}\ {\rm cm^2 \ s^{-1}}$ may, indeed, be
larger than $r_{BH} v \sim 10^{18} - 10^{19}\ {\rm cm^2 \ s^{-1}}$.
The ion-neutral bias is particularly applicable in dense clouds
$n_4\ \sgreat \ 1$ or, more precisely, when
$n_4^{1/3} v_{40}^{-1} \ \sgreat \ 70$,
otherwise the accretion radius is well within the ionization
sphere and the inflow is ionized in the first place.

Gamma-ray observations show a sharp concentration of high energy radiation
in a very thin Galactic disk, with evidence for
imprints of the Galactic arms. If this emission comes from decaying pions,
originating in cosmic ray interactions with the ambient gas
(Stecher \& Stecker, 1970), it requires
the cosmic ray concentration to follow the high energy photons.
In our model this
is fulfilled since the dense ISM clouds
are also concentrated in the spiral arms,
generally close to the galactic equator.

What is the electromagnetic imprint of isolated NSs?
The soft X-ray luminosity from the shock region
is a fraction of $ \propto L_{cr}(1 - f_{cr})$, and
the expected spectrum is nonthermal.
{}From equations 13 and 14 one
can predict a power law of spectral index $(\alpha - 1)/2 \sim 1/2$.
Other emission mechanisms, involved with
heat loss of the star itself would
basically produce thermal emission, but a magnetized
atmosphere could affect the radiation
transfer and modify the blackbody curve.
A crucial point to note is that
the Galactic diffuse X-ray background hardly restricts the
cosmic - ray luminosity from isolated NSs.
The integrated flux from isolated NSs was found to contribute no more than
a fraction of a per cent to the soft X-ray background (BM93).
The strongest accretors are located in dense clouds, and so even
if they emit strong soft X-ray radiation, it would be greatly absorbed.
Using a photoionization  cross section
$\sim 10^{-22} (E/{\rm KeV})^{-3} \ {\rm cm^2 \ (H-atom)^{-1}}$,
the cloud optical depth
$\tau_x \sim 10^2  (E/{\rm KeV})^{-3} n_4
(R/10^{20}{\rm cm})$,
where $R$ is the cloud radius.
Solitary isolated NSs, moreover,
may be the origin of the enhanced X-ray emission within
$\pm 40^0$ longitude of the Galactic center,
as observed by {\it EXOSAT} (BM93, Maoz \& Grindlay 1994).

Although
isolated NSs are the majority of the NS population,
they have so far received only limited attention compared with that given
to the brighter ones, the pulsars and the (NS) X-Ray sources.
Only
a small fraction of the Galactic isolated NSs can be observed directly.
Following the early suggestion that nearby
isolated NSs might be detectable due to radiation released from the accretion
of the ISM
(Ostriker, Rees \& Silk 1970),
Madau \& Blase (1994)
argue that the $\sim 100$
sources detected in the EUV - soft X-ray range
the {\it ROSAT - WPC} and {\it EUVE}
all - sky surveys,
and which have no previous identification in the optical or other energies,
are the natural candidates for isolated NSs.
Recent observations at the Wise Observatory of the fields of $\sim 15$
of these sources show that indeed in many cases the sources
are likely hot or peculiar,
and unlikely to be white dwarfs or late type
flare stars.
Simultaneous observations planned from
a number of experiments aboard the
{\it Spectrum X Gamma} observatory, in UV ({\it TAUVEX}),
EUV ({\it EUVITA}),
and X-Ray bands ({\it SODART, Jet-X, MART}),
will be extremely useful for the identification of such objects.
The predicted characteristics to be studied in such UV - X-ray observations
include spectral index, the presence of synchrotron absorption lines,
temporal modulation, proper motion, and HII cometary - like regions.
Detection of isolated NSs would be
an important step towards our understanding of this Galactic component.
The model presented here can be constrained
by observations of the total isolated NS population, their
spatial and velocity distributions, and the fraction of them in dense clouds.

\bigskip
\rf
{\bf ACKNOWLEDGMENTS}
\medskip
\rf
Astronomy at the Wise Observatory is supported by grants from the
Ministry of Science and Technology and from the Israel Academy of Science.
The author thanks the referee Luke O.\'C. Drury for his comments
and help in improving this paper, and
Peter Biermann, David Eichler,
Jonathan Katz, Amir Levinson, Dani Maoz and
Vladimir Usov for discussions.

\vfill\break
\rf
{\bf REFERENCES}
\medskip
\rf
Arnett,W.D., Schramm,D.N.,\& Truran, J.W. 1989, ApJ, 339, L25
\rf
Arons, J. \&  Tavani, M., 1994, ApJ Suppl, 90, 797
\rf
Axford, W.I., 1994, ApJ Suppl. 90, 937
\rf
Bell, A.R., 1978, MNRAS, 182, 147
\rf
Blaes, O. \& Madau, P. 1993, ApJ, 403, 690 (BM93)
\rf
Blandford, R.D. \& Eichler, D. 1987, Phys. Rep. 154, 1 (BE87)
\rf
Blandford, R.D. \& Ostriker, J.P., 1980, ApJ, 237, 793
\rf
Cesarsky, C.J. 1987, in {\it High Energy Phenomena Around Collapsed Stars},

ed: Pacini, A., 331, Reidel
\rf
Cowsik, R. \& Lee, M. A., 1982, Proc. R. Soc. Lond. A 383, 409
\rf
Draine, B.T. \& McKee, C. F. 1993, Ann. Rev. Ast. Ast., 31, 373
\rf
Drury, L.O.\'C. 1983, Rep. Prog. Phys. 46, 973
\rf
Ellison, D.C., \& Eichler, D. 1985, Phys. Rev. Lett., 55 No. 24, 2735
\rf
Geiss, J., 1982, Spa. Sci. Rev., 33, 207
\rf
Illarionov, A.F. \& Sunyaev, R.A. 1975, A\&A, 39, 185
\rf
Kazanas, D. \& Ellison, D.C., 1986, Nature, 319, 380
\rf
Madau, P. \& Blaes, O. 1994, ApJ (Preprint)
\rf
Meyer, J.P. 1985, ApJ Suppl. 57, 151 and 173
\rf
Nelson, W.R., Wang, J.C.L., Salpeter, E.E., \& Wasserman, I. 1994 ApJ Letter
\rf
Ostriker, J.P., Rees, M.J. \& Silk, J. 1970, Astrophys. Lett., 6, 179
\rf
Phinney, E.S. \& Kulkarni, S.R., 1994, Ann. Rev. Ast. Ast., 32, 591
\rf
Ruffert, M., 1994, preprint MPA 789, submitted to AA
\rf
Shapiro, S.L. \& Teukolsky, S.A., 1983, {\it Black Holes, White Dwarfs, and
Neutron Stars},

John Wiley \& Sons
\rf
Shemi, A., 1995, in: Conference Proceedings of {\it Shocks in Astrophysics},
UMIST January 1995
\rf
Silberberg, R., Tsao, C.H., Shapiro, M.M. \& Biermann, P.L., 1990, ApJ, 363,
265
\rf
Srinivasan, G. 1989, A\&A Rev. 1, 209
\rf
Stecher, T.P. \& Stecker, F.W., 1970, Nat, 226, 1234
\rf
Turolla, R., Zampieri, I., Colpi, M. \& Treves, A. 1994, ApJ, 426, L35
\rf
V\"olk, H.J. \& Biermann, P.L. 1988, ApJ, L65

\end